Statistical Approach to Radiation Processes with Heavy Atoms in Plasmas

A.V. Demura[*], M.B. Kadomtsev, V.S. Lisitsa, V.A. Shurygin

NRC "Kurchatov institute", Academician Kurchatov square 1,
Moscow 123182

Abstract

The new statistical approach for calculation of radiation processes with heavy multielectron ions in plasma is developed. The method consists in consideration of atomic structure as a condensed medium, characterized by the spectrum of elementary excitations with plasma frequency, determined by local atomic electron density. For instance, the radiation losses in this model are due to excitation of plasma type oscillations in atom under its collisions with plasma electrons and have a universal statistical representation for all sorts of multielectron ions. The calculations of radiation losses on tungsten ions are performed in the wide range of plasma temperature variation, typical for physics of high temperature plasma with magnetic confinement. It is shown that the universal statistical approach results are within the scattering of current numerical codes data. The proposed statistical method for description of complex atoms collective excitations for calculations of plasma radiation losses is of general physical interest and allows to obtain the necessary data more faster with the lesser computational resources.

The calculations of radiation losses of heavy atoms in plasmas acquired particular interest due to implementation of tungsten in construction elements of current thermonuclear installations [1]. The diapason of temperature being of interest for calculations of radiation losses turns out to be extremely large – from several electron-volts in the SOL and divertor plasmas up to 40 keV in the central regions [1]. The energy structure of multielectron states of tungsten ions is very complex in all temperature diapason that requires cumbersome and time-consuming quantum mechanical calculations of atomic structure, as well as elementary processes, responsible for population of atomic levels. As under the calculations of rate coefficients the additional approximations are used, that work well only in the limited temperature diapason there are significant deviations between results of complex detailed codes [2-3]. Therefore for the description of heavy ions structure it is natural to use general statistical methods [4-5], that allow to retrieve scalings of radiation processes in all range of temperatures. In this approach the atomic spectra could be represented as collective excitations of condensed medium [4]. The proposed in present work implementation of statistical models [6-7] allows to elaborate the universal statistical approach for analysis of radiation losses and the simple method of their calculation.

The statistical models are based on the notion of collective oscillations of atomic electrons plasma bunch. For the description of this oscillations in the Brandt-Lundquist model [6] the approximation of local plasma frequency (LPF) is used, connected with the local electron atomic density. In the Vinogradov-Tolsctikhin paper [7] on the basis of solution of the kinetic Vlasov equation it is shown that the approach [6] does not take into account polarization field, inducted by the external atomic perturbation. The discrepancy between both approaches for the photoionization cross sections, however, is within the accuracy of radiation losses calculations for heavy atoms (ions), and below the both indicated approximations are used.

The representation of atomic structure as a system of oscillators, being excited by collisions with external plasma electrons, is in the background of the present approach. The interaction of plasma electrons with atoms is considered in the Fermi approximation of equivalent photons [8], where the electric field of equivalent photon flux is determined by the

---

[*] e-mail: demura45@gmail.com, mkadomtsev@mail.ru, vlisitsa@yandex.ru, va-sh@yandex.ru

Fourier-expansion of the electric field of electron, moving along the classical trajectory in the field of atom being excited. In this formulation the excitation of bound electrons in multielectron ion is expressed in terms of the photoabsorption cross section, for which in its turn the mentioned above statistical models of heavy atoms could be used [6-7].

One of the significant general property of the atomic systems is the evident relation between the effective atomic oscillator strengths $f_{ij}$ and the atomic or ionic electron density distribution n(r)

$$f_{if} = 4\pi \cdot n(r) \cdot r^2 \cdot dr, \tag{1}$$

which ensures the fulfillment of the well known sum rule [9].

Within the local plasma frequency approximation [6] the complex ion is represented as a system of equivalent oscillators, whose frequencies are determined by the values of local plasma frequency $\omega_p(r)$ along with the well known formula (atomic units)

$$\omega_p(r) = \sqrt{4\pi n(r)} \tag{2}$$

As could be shown the conditions of coronal model, in which the radiation losses are determined totally by collisional excitation rates for ions, are fulfilled for applications considered below. Therefore within mentioned approximations the radiation losses of energy per an electron and per a given ion with the charge $Z_i$ of heavy impurity (ie with given value of $q = Z_i/Z$) takes the universal form, expressed in terms of photoexcitation rates of atoms in the field of equivalent photons

$$Q_{abs} / n_e = Z \cdot \left(\frac{2Ry \cdot \omega_a}{a_0^2}\right) \cdot \int_0^{I_Z/Z\hbar\omega_a} d(\omega/\omega_a Z) \cdot \sigma_{photo}(\omega/\omega_a Z) \cdot \left\{ \frac{<dI^{(Coulomb)}[Z(\omega/\omega_a Z)]>_E}{n_e \left(\frac{2Ry \cdot \omega_a}{a_0^2}\right) d(\omega/\omega_a)} \right\} =$$

$$= (a_0^3) \cdot (2Ry \cdot \omega_a) \cdot \left(\frac{c\hbar}{e^2} \cdot \sqrt{\frac{1}{6\pi}}\right) \cdot \sqrt{\frac{2Ry}{T}} \cdot (Z) \int_0^{I_Z/Z \cdot 2Ry} ds \cdot [\sigma_{photo}(s)/a_0^2] \cdot \tag{3}$$

$$\cdot \int_{Z\left(\frac{2Ry}{T}\right) \cdot s}^{\infty} du \cdot e^{-u} \cdot g\left(\left[\frac{Z_{eff} Z}{2\sqrt{2}} \cdot \left(\frac{2Ry}{T}\right)^{3/2}\right] \cdot s \cdot u^{-3/2}\right),$$

where $a_0$ is the Bohr radius, $\omega_a = 2Ry/\hbar$, c is the speed of light, e is the electron charge, $I_z$ is the ionization potential of the given ion, $\sigma_{photo}(x)$ is the photoabsorption cross section of the given ion, g(z) is the so called Gaunt factor, describing trajectory curvature in the Coulomb field, $<dI^{(Coulomb)}[Z(\omega/\omega_a Z)]>_E$ is the intensity of equivalent photon flux with the circular frequency $\omega$ per unit circular frequency interval, averaged over the energies of electron projectiles E in assumption of the Coulomb trajectories of electrons being scattered by the target. Here the motion of incident electron in the Thomas-Fermi potential is approximated by its motion in the Coulomb potential with some effective charge $Z_{eff}$. This allows to express the intensity of equivalent photons via the Gaunt-factor in the Coulomb potential. Concurrently in the local plasma frequency model the effective charge is determined from the condition of equality of the Thomas-Fermi potential and the Coulomb potential in the point $r_\omega = r_{TF} x_\omega$ ($r_{TF}$ is the Thomas-Fermi radius), corresponding to the condition of resonance of the absorbed $\omega$ and plasma $\omega_p = \sqrt{4\pi n(r_\omega)} = \omega$ frequencies. Taking into account the relation between density and potential in the Thomas-Fermi model, it is possible to obtain the expression for effective charge

$$Z_{eff} = Z \left\{ \chi(r_\omega) + \frac{q r_\omega}{r_0} \right\}, \tag{4}$$

where qZ is the ion charge, $r_0$ is the distance from the nuclei where the electron density of the ion with the given charge go to zero in the Thomas-Fermi model, $\chi(x,q)$ is the known function, describing the behavior of potential and density in the Thomas-Fermi model for the ion with the nuclei charge Z and stripping q [5]. The $Z_{eff}$ value changes smoothly from the ion charge qZ at small frequencies to the charge of nuclei Z at large frequencies. This variation for the tungsten ion with the charge $Z_i = qZ = 22$ is shown in the Fig. 1.

The integration over frequencies is performed up to the ionization potential of the given ion that corresponds to taking into account only the bound states. The integration over energies of incident electron goes from the equivalent photon frequency, which corresponds to the excitation thresholds of atomic transitions in the statistical model. The photoabsorption cross sections are taken below along with models [6,7]. The value of Gaunt-factor in the Coulomb approximation [10] is equal to

$$g(v) = \frac{\pi\sqrt{3}}{4}\left\{iv\, H_{iv}^{(1)\prime}(iv)H_{iv}^{(1)}(iv)\right\} \approx \frac{\sqrt{6}}{\pi}\ln\left[\left(\frac{2}{\gamma v}\right)^{1/\sqrt{2}} + e^{\pi/\sqrt{6}}\right], \quad (5)$$

where $H_p^{(1)}(z), H_p^{(1)\prime}(z)$ are Hankel function and its first derivative over argument, $\gamma \approx 1.78$ – is the Euler's constant..

If to neglect the Gaunt-factor variation, then

$$\frac{Q_{abs}}{n_e} = (a_0^3)\cdot(2Ry\cdot\omega_a)\cdot\left(\frac{c\hbar}{e^2}\cdot\sqrt{\frac{1}{6\pi}}\right)\cdot\sqrt{\frac{2Ry}{T}}\cdot(Z)\int_0^{I_Z/Z\cdot 2Ry} ds\cdot[\sigma_{photo}(s)/a_0^2]\cdot\exp\left[-Z\cdot\left(\frac{2Ry}{T}\right)\cdot s\right] \quad (6)$$

For calculation of the total radiation losses on all ions at the given temperature it is necessary to sum the expressions (3,6) using the corresponding equilibrium ionization distribution [11].
In the specific calculations the two statistical models were used: the model of local plasma frequency (LPF) [6] and the electrodynamical model (EM) [7].

In the LPF model [6] the photoabsorption cross section is expressed in the form:

$$\sigma_{abs}(\omega) = \frac{2\pi^2 e^2}{mc}\int d^3r\, n(r)\delta(\omega-\omega_p(r)) = \frac{2\pi^2 e^2}{mc}\cdot 4\pi\left(r_\omega^2\cdot\frac{n(r_\omega)}{\left|\frac{d\omega_p(r)}{dr}\right|_{r=r_\omega}}\right), \quad (7)$$

where m is the electron mass.

If to use additionally the Thomas-Fermi model for the atomic electron density distribution, then equations (6-7) are transformed to

$$Q_{abs}/n_e = (a_0^3)\cdot(2Ry\cdot\omega_a)\cdot\left(\frac{3\pi^4}{16}\cdot\sqrt{\frac{1}{6\pi}}\right)\cdot\sqrt{\frac{2Ry}{T}}\cdot(Z)\int_0^{I_Z/Z\cdot 2Ry} ds\cdot s\cdot\frac{x_s^2\frac{\chi(x_s,q)}{|\chi'(x_s,q)|}}{\left|1-\frac{\chi(x_s,q)}{x_s\chi'(x_s,q)}\right|}\cdot$$

$$\cdot\int_{Z\cdot\left(\frac{2Ry}{T}\right)\cdot s}^{\infty} du\cdot e^{-u}\cdot g\left(\left[\frac{Z_{eff}Z}{2\sqrt{2}}\cdot\left(\frac{2Ry}{T}\right)^{3/2}\right]\cdot s\cdot u^{-3/2}\right), \quad (8)$$

and for g(z)=1

$$\frac{Q_{abs}}{n_e} = (a_0^3)(2Ry\cdot\omega_a)\cdot\left(\frac{3\pi^4}{16}\cdot\sqrt{\frac{1}{6\pi}}\right)\cdot\sqrt{\frac{2Ry}{T}}\cdot Z\int_0^{I_Z/Z\cdot 2Ry} ds\cdot s\cdot\frac{x_s^2\frac{\chi(x_s,q)}{|\chi'(x_s,q)|}}{\left|1-\frac{\chi(x_s,q)}{x_s\chi'(x_s,q)}\right|}\cdot\exp\left[-Z\left(\frac{2Ry}{T}\right)s\right]. \quad (9)$$

In the case of EM model the numerical values of corresponding photoabsorption cross sections from [7] were substituted in (6).

The results of calculations are presented in the Fig. 2. Here only the contribution in the radiation losses due to the collisional excitation of atoms by plasma electrons, described in the Fermi approximation of equivalent photons is presented. The contribution due to the radiative recombination (RR) turns out to be small under domination over the RR the collisional excitation (CE) of the ion core, possessing sufficiently large number of electrons, even at high temperature values. What concerns dielectronic recombination (DR) its contribution also was calculated on the basis of the known Burgess's formula [12] and the statistical approach, that allows to express the oscillator strengths and the transition energies in this formula via the atomic electron density. It was found, that the DR contribution amounts at most 10% in all the range of temperature variation, and therefore it could be neglected within the statistical model accuracy itself. The results of losses calculations by the known numerical codes [2-3, 13-17] are presented in the Fig.2 as well. It is seen, that the discrepancy between the results of these numerical calculations turns out to be of the same order as for the statistical models. The largest difference with the numerical calculations is observed in the region of small temperatures, where the excitation of the outer ions shell becomes essential, for which the implementation of the statistical model starts already to be problematic. At the same time the interesting circumstance is the sufficiently good conformity of the results of detailed numerical calculations [17] and those due to the statistical model at lowest temperatures 1-2 eV.

Let us compare the results based on photoexcitation cross sections and equivalent photon method with the standard electron excitation theory, both modified by the additional applications of statistical atomic model. Indeed, the excitation rates for dipole allowed transitions can be expressed in terms of the excitation energies $\Delta E_{ij}$ and corresponding oscillator strengths $f_{ij}$ [12]

$$\langle v\sigma_{ij} \rangle (cm^3 \sec^{-1}) = 3,2 \cdot 10^{-7} \cdot f_{ij} \cdot \left(\frac{Ry}{\Delta E_{ij}}\right)^{3/2} \beta_{ij}^{1/2} e^{-\beta_{ij}} p(\beta_{ij}) \qquad (10)$$

where $\beta_{ij} = \Delta E_{ij}/T$, $p(\beta_{ij})$ is the tabulated function (see [12]).

Then after the multiplication of excitation rates by the electron transitions energies and summation over all possible transitions $i \to j$ in the given ion with $q_k = k/Z$, and next the substitution in the resulting expression the atomic parameters in terms of statistical electron density according to (1,2) one arrives to the simple formula for partial radiative energy losses

$$\frac{Q_{EXC,k}(Wm^3)}{4,9 \times 10^{-31}} = Z\sqrt{\frac{2Ry}{T}} \cdot \int_0^{x_0(q_k)} dx \cdot x^2 \cdot \left(\frac{\chi(x,q_k)}{x}\right)^{3/2} \cdot p\left(-1,2 \cdot Z\sqrt{\frac{2Ry}{T}} \left(\frac{\chi(x,q_k)}{x}\right)^{3/4}\right)$$
$$\cdot \exp\left[-1,2 \cdot Z\sqrt{\frac{2Ry}{T}} \left(\frac{\chi(x,q_k)}{x}\right)^{3/4}\right]. \qquad (11)$$

The information on atomic structure is contained in (11) via the atomic density distribution for specific $k$ ion with the parameter $q_k = k/Z$. In the averaged ion approximation (AIA) the parameter $<q> = \sum_{k=0}^{Z} k P_k / Z$ value, averaged over the ionization equilibrium distribution $P_k(T)$, is used in (11) instead of $q_k$ to obtain the total radiation losses $Q_{EXC}$ due to the electronic excitations of ions in plasmas.

The comparison of the electromagnetic (EM) and local plasma frequency (LPF) approximations with the electron excitation theory, modified by Thomas-Fermi and Brandt-Lundquist models for ions and marked by EXC, is presented in the Fig.3. One can see a good correspondence between the statistical models, especially between EXC and LPF results. Obviously this is due to known good correspondence between classical and quantum electron excitation theories for dipole allowed transitions (see [12]).

Thus the results of this work are the first experience of application of the statistical models for the calculation of plasma radiation losses on heavy impurities. It is shown, that the results of universal statistical approach are within the data scattering of known numerical codes [2-3, 13-17], using different approximations for the calculation of atomic structure and the electronic excitation cross sections of complex ions. The further development of the method might consists in taking into account the shell effects [18], becoming apparent at the periphery of atomic electron density distribution.

The developed in the paper universal statistical approach for analysis and calculation of the radiation losses proceeds from the *ab initio* principles of statistical theory, where all atomic characteristics are expressed via the distribution of atomic electron density, whilst the sum rule (the conservation of the sum of transitions oscillator strengths) is the essential condition of consistent realization of the general statistical approach idea.

The employed in this work method of collective excitations of atoms for the calculations of plasma radiation losses is of general physical interest. Its significance for applied science consists in the elaboration of statistical method for calculation of plasma radiation losses on heavy ions, which allows to obtain necessary data significantly faster and with the lesser consumption of computational resources.

The work was partially supported by the RFBR grant № 13-02-00812 and by the НШ--3328.2014.2 grant of RF president for the state support of RF leading scientific schools.

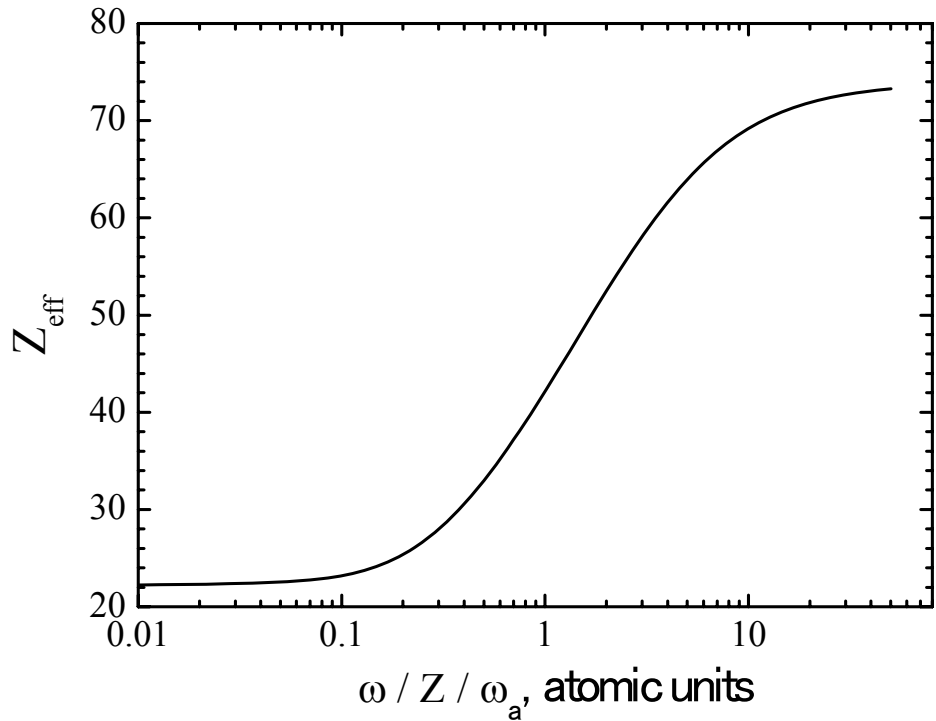

Fig. 1. Effective charge in Thomas-Fermi model along with Eq (4) versus reduced circular frequency for ion $W^{22+}$.

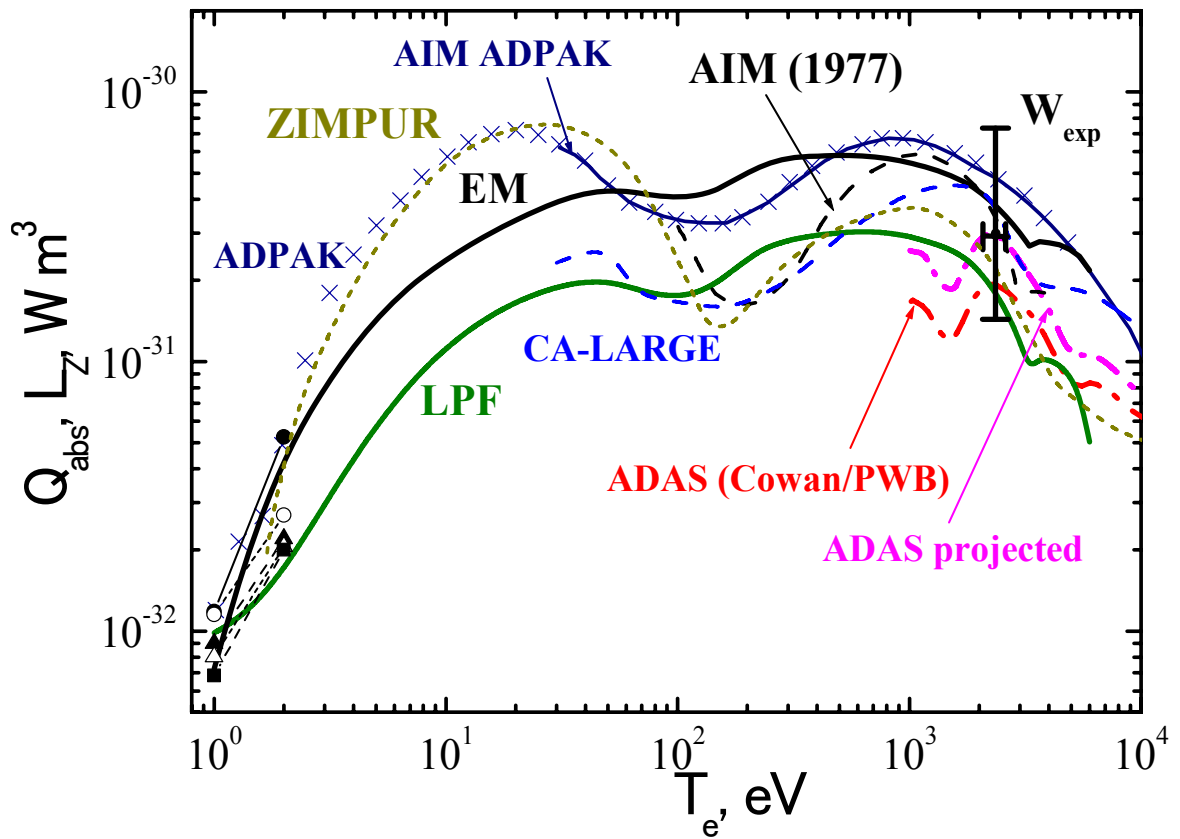

Fig. 2. Comparison of radiation losses on tungsten impurity within universal statistical approach (EM – electromagnetic method, LPF – method of local plasma frequency) with results of known codes versus plasma temperature: ADPAK - [2]; AIM ADPAK [3]; AIM – averaged ion model [13]; ADAS projected -[3,14]; ADAS COWAN/PWB - [3,14]; ZIMPUR [15]; CA-LARGE - [16]; dark circles - ADPAK, light circles- CFG-AVE, dark triangles - FS-NOCI, light triangles - FS-CI, dark squares - FS-FOM data of radiative-collisional models from [17]; $W_{exp}$ – experimental estimate of radiation losses value [16]

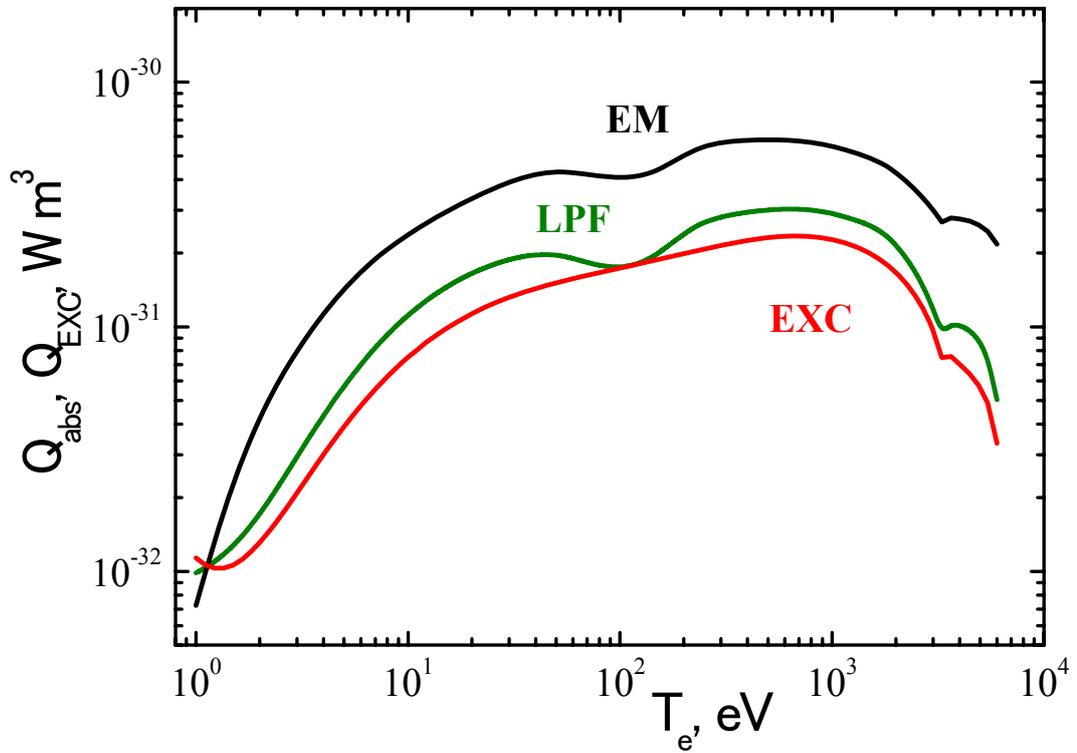

Fig.3. Comparison of radiation losses on tungsten impurity within the universal statistical approach (EM – electromagnetic method, LPF – method of local plasma frequency) with the electron excitation theory results [12] in AIA marked as EXC.